\documentclass[traditabstract]{aa}
\usepackage{txfonts}
\usepackage{graphicx}
\pdfoutput=1

\usepackage{color}
\usepackage{subfig}

\newcommand{\be}{\begin{equation}}
\newcommand{\ee}{\end{equation}}
\newcommand{\ud}{{\rm d}}

\begin{document}

\title{Adaptable Radiative Transfer Innovations for Submillimetre Telescopes (ARTIST)}

\subtitle{Dust polarisation module (DustPol)}

\author{M. Padovani\inst{1}
          \and
          C. Brinch\inst{2}
          \and
          J.~M. Girart\inst{1}
          \and
          J.~K. J{\o}rgensen\inst{2}
          \and
          P. Frau\inst{1}
          \and
          P. Hennebelle\inst{3}
          \and
          R. Kuiper\inst{4}
          \and
          W.~H.~T. Vlemmings\inst{5}
          \and
          F. Bertoldi\inst{6}
          \and
          M. Hogerheijde\inst{7}
          \and
          A. Juhasz\inst{7}
          \and
          R. Schaaf\inst{6}
          }

\authorrunning{M. Padovani et al.}
\titlerunning{ARTIST - Dust polarisation module (DustPol)}

    \institute{Institut de Ci\`encies de l'Espai (CSIC--IEEC), 
    Campus UAB, Facultat de Ci\`encies, Torre C5p, 
    08193 Bellaterra, Catalunya, Spain\\
    \email{[padovani,girart,frau] @ice.cat}
    \and
    Centre for Star and Planet Formation and Niels Bohr Institute, University of 		    
    Copenhagen, Juliane Maries Vej 30, 2100 Copenhagen {\O.}, Denmark\\
    \email{[brinch,jeskj] @nbi.dk}
    \and
    Laboratoire de radioastronomie, UMR 8112 du CNRS, \'Ecole normale sup\'erieure 
    et Observatoire de Paris, 24 rue Lhomond, 75231 Paris Cedex 05, France\\
    \email{patrick.hennebelle@lra.ens.fr}
    \and
    Jet Propulsion Laboratory, 4800 Oak Grove Drive,
	Pasadena, CA 91109, USA\\
	\email{rolf.kuiper@jpl.nasa.gov}
	\and
	Department of Earth and Space Sciences, Chalmers University of Technology, 
	Onsala Space Observatory, 439 92 Onsala, Sweden\\
	\email{wouter.vlemmings@chalmers.se}
    \and
    Argenlander Institute for Astronomy, University of Bonn, 
    Auf dem H\"ugel 71, 53121 Bonn, Germany\\
    \email{[bertoldi,rschaaf] @astro.uni-bonn.de}
    \and
    Leiden Observatory, Leiden University, PO Box 9513, 
    2300 RA Leiden, The Netherlands\\
    \email{[michiel,juhasz] @strw.leidenuniv.nl}
    }
\date{}

 
 \abstract
   {We present a new publicly available
   tool ({\tt DustPol}) aimed to model the polarised thermal dust
   emission. The module {\tt DustPol}, which is publicly available,
   is part of the ARTIST (Adaptable Radiative Transfer Innovations for Submillimetre 	
   Telescopes) package, which also offers tools for modelling the polarisation
   of line emission together with a model library and a Python-based user interface.
   {\tt DustPol} can easily manage analytical as well as pre-gridded models to generate
   synthetic maps of the Stokes $I$, $Q$, and $U$ parameters. These maps are stored 
   in FITS format which is straightforwardly read by the data reduction
   software used, e.g., by the Atacama Large Millimeter Array (ALMA). 
   This turns
   {\tt DustPol} into a
   powerful engine for the prediction of 
   the expected polarisation features of a source
   observed with ALMA or the Planck satellite as well as for the 
   interpretation of existing submillimetre observations
   obtained with other telescopes. 
   {\tt DustPol} allows the parameterisation of the maximum degree of polarisation and
   we find that, in a prestellar core, if there is depolarisation, 
   this effect should happen at densities of $10^6$~cm$^{-3}$ or larger.
   We compare a model generated by 
   {\tt DustPol} with the observational polarisation data of the low-mass Class~0
   object NGC~1333 IRAS~4A, finding that the total and the polarised emission
   are consistent.}


   \maketitle
%


\section{Introduction}
The importance of magnetic fields in regulating the physics of star formation is
still a matter of debate. The main
schools of thought invoke either the leakage of magnetic support through
ambipolar diffusion (Mestel \& Spitzer~\cite{ms56}) or the supersonic 
turbulence which compresses the gas at the interfaces between converging
flows (Elmegreen~\cite{e93}).
While turbulent energy is required to sustain a cloud over its lifetime
(Gammie \& Ostriker~\cite{go96}), ohmic dissipation time of magnetic fields is much larger than the collapse time, assuming a typical ionisation fraction of
$\sim10^{-8}-10^{-7}$ and density of $\sim10^{5}$~cm$^{-3}$
(Pinto \& Galli~\cite{pg08}).
Collapse time scale for a subcritical core is a few million to 10$^7$~yr
assuming the 
ambipolar diffusion to be the dominant process (Mouschovias et al.~\cite{mt06}).
Conversely, if collapse is
controlled by turbulence, the time scale is shorter, of the order of 
10$^{6}$ yr (Mac Low \& Klessen~\cite{mk04}). Cores can also be forced into collapse
because of external events such as collision between clouds as well as
supernova shock waves with a time scale as short as 10$^5$ yr 
(Boss et al.~\cite{bk10}). All these different mechanisms most likely take part
in cloud collapse, but the role of magnetic fields cannot be ruled out from
star formation theories, even when a core is supercritical (Basu~\cite{b97}).

From an observational point of view, it is possible to gather information about
the strength and the geometry of magnetic fields through measurements
of the Zeeman splitting of hyperfine molecular transitions as well as measurements of
polarisation of thermal dust emission in the (sub)millimetre wavelength range.
Observations of the Zeeman effect provide information about the line of sight component
of the magnetic field and have been carried out using atoms and molecules which show
large magnetic moments
(e.g. Crutcher et al.~\cite{ct96}, Falgarone et al.~\cite{ft08}).
By observing maser emission it is possible to detect both 
circular and linear polarisation, making it possible to derive the full 3D
magnetic field (Vlemmings et al.~\cite{vh11}, Surcis et al.~\cite{sv11a}, \cite{sv11b}).
On the other hand, polarisation of the thermal dust emission probes the plane of the sky
component of magnetic fields, providing information about 
the field geometry in molecular
clouds and filaments associated with star formation regions
(e.g. Novak et al.~\cite{ng89}, Matthews et al.~\cite{mw01},
Davidson et al.~\cite{dn11}). (Sub)millimetre wavelength 
interferometric observations, first with the 
Berkeley-Illinois-Maryland Association array, BIMA
(e.g. Girart et al.~\cite{gc99}, Lai et al.~\cite{lg03}), and nowadays with 
the Submillimeter Array, SMA
(Girart et al.~\cite{gb09}, Rao et al.~\cite{rg09}, Tang et al.~\cite{th09}) have
proven useful to study the magnetic field properties at high angular resolution,
probing scales of a few hundreds AU
at which the dynamical collapse of the star forming cores is
occurring.

The Atacama Large Millimeter Array (ALMA) is the largest interferometer and the largest
ground based project in (sub)millimetre
astronomy. It provides a considerable enhancement in
resolution, sensitivity and imaging with respect to previous interferometers 
operating at these wavelengths. Since dust and a multitude of spectral lines
can be observed at the same time, ALMA allows an in-depth investigation of the
physical and chemical properties of e.g. embedded protostars,
zooming in to AU scales of nearby star-forming regions.
ALMA will also open new possibilities for studying magnetic fields. In fact, due to 
the receiver capabilities, full polarisation calibration and imaging will be the norm
rather than the exception.
For these reasons, providing ALMA users with self-consistent tools for polarisation modelling will be very important. The 
publicly available Adaptable Radiative Transfer Innovations for 
Submillimetre Telescopes\footnote{\tt http://youngstars.nbi.dk/artist} 
(ARTIST) 
package
(J\o rgensen et al. {\em in prep.}) 
provides a direct link between the theoretical predictions and the
quantitative constraints from the submillimetre observations, being able to model
the full multi-dimensional structure of, e.g., a low-mass protostar, including its envelope, disk, outflow, and magnetic field.

The model suite is based on LIME\footnote{\tt http://www.nbi.dk/$\thicksim$brinch/lime.html} (Line Modelling Engine, Brinch \& Hogerheijde~\cite{bh10}), 
a non-LTE radiative transfer 
code using adaptive gridding  in three dimensions
that allows simulations of sources with arbitrary 
multi-dimensional structures ensuring a rapid convergence. 
In addition, ARTIST 
offers unique tools for modelling the polarisation of dust and line
emission and an extensive library of theoretical (semi-analytical) models.
All these packages are connected by a
comprehensive Python-based graphical user interface with 
direct link to, e.g., ALMA data reduction software 
(Common Astronomy Software Applications package, 
CASA\footnote{\tt http://casa.nrao.edu}).

In this paper, we describe the dust polarisation module of ARTIST:
in Sect.~\ref{physicalproblem} we briefly describe the  
polarisation of stellar radiation due to dust grains,
in Sect.~\ref{tools} we explain the features of the module for analysing dust
polarisation emission, and in Sect.~\ref{applications} we demonstrate the capabilities
of the routines to handle analytical models, numerical simulations, 
and providing a comparison to observations.
Finally, in Sect.~\ref{conclusions} we summarise our results and conclusions.


\section{Origin of dust polarisation}\label{physicalproblem}
Dust grain alignment provides a natural explanation for the origin
of polarisation in the ISM. However, what mechanisms are responsible for grain alignment and what role they play in different environments remain poorly understood
(Lazarian~\cite{l03}, \cite{l07}). The main
explanations are supplied by paramagnetic (Davis \& 
Greenstein~\cite{dg51}) 
and mechanical alignment (Gold~\cite{g51}), but other mechanisms have been proposed 
as well (Martin~\cite{m71}, Purcell~\cite{p79}, Draine \&
Weingartner~\cite{dw96}).
Recent studies for example show
that radiative torques, which are the results of the interaction of irregular grains with a 
flow of photons,
can effectively align the dust grains 
(Dolginov \& Mytrophanov~\cite{dm76}, Lazarian~\cite{l07}).

Grains are most likely non-spherical 
since they are formed by stochastic growth processes leading to irregular shapes
(Fogel \& Leung~\cite{fl98}). Unpolarised starlight passing
through an aligned-grain set will acquire polarisation due to the different
grain extinction parallel and perpendicular to the alignment direction, since
they absorb more light along their longer direction.
This depends on the fact that extinctions are proportional to the parallel and
perpendicular grain cross section (see Lee \& Draine~\cite{ld85} and
Efroimsky~\cite{e02} for a detailed discussion for spheroidal grains and a more
general case, respectively).

Conversely, in the submillimetre wavelength regime, 
the electric-field vector of the thermal continuum 
emission emerging from rotating aligned grains has its maximum value in the plane 
containing the longer grain axis. The partially linearly polarised vectors 
are parallel to 
the mean grain orientation, therefore perpendicular to the magnetic field as
projected on the plane of the sky. Consequently, from the observation of
the polarised dust emission it is possible to model the morphology of magnetic
fields in the plane of the sky, the degree of polarisation depending 
on various contributions as the
grain cross sections, the temperature, the composition, and the density.



\section{Dust polarisation tools}\label{tools}
With the dust polarisation module ({\tt DustPol}) of ARTIST we aim
to model the polarised dust
continuum emission. 
The {\tt DustPol} module is an extension to the LIME radiative transfer code.
LIME is
a non-LTE molecular excitation and line radiation transfer 
code that is mainly used to calculate line profiles in the
far-infrared and submillimetre regimes. However, LIME 
can also ray-trace given density and temperature profiles and from that estimate the emerging
continuum flux. It is
this feature that we make use of in the {\tt DustPol} module presented in this paper. 
At the
relatively long wavelengths at which LIME operates, 
scattering is 
negligible and therefore the problem of making an image of 
the continuum model is reduced to integrating the source
function
\begin{equation}
{I_\lambda = \int _0 ^\infty B_\lambda(T_{\rm dust})\ e^{-\tau_\lambda}\,\ud\tau_\lambda}
\end{equation}
along lines of sight through the model.
$B_{\lambda}(T_{\rm dust})$ is the Planck
function evaluated at the dust temperature $T_{\rm dust}$ and 
$\ud\tau_\lambda = \kappa_\lambda\ \!\rho_{\rm dust}\ \!\ud s$, where 
$\kappa_\lambda$ is the wavelength dependent dust 
opacity and $\rho_{\rm dust}$ is the dust mass density.
The main benefit of
using the LIME code for our application is its 
ability to automatically map a 3D model onto a computational
grid. The grid points are placed at random in 3D space with 
probability of a point being placed at a given location 
weighted by a user-selected function, typically
the underlying model dust density distribution. This 
results in a higher spatial grid point density in high dust
density regions and vice versa for low density regions, which
means that the expectation distance between neighbour points 
scales with the photon mean free path.

Once the grid points have been placed, the Voronoi diagram of the distribution and the 
topological equivalent, the Delaunay triangulation, is calculated
(Ritzerveld \& Icke~\cite{ri06}). A Voronoi region 
associated with a certain grid point is defined as the set containing points that
is located 
closer to the generating grid point than to any other grid point. The result is a 
tessellation of space consisting of random polyhedra. All physical properties are taken 
to be constant over each polyhedra (or Voronoi cell). If the model resolution turns out 
insufficient in specific regions, more grid points can be added to these 
and the Voronoi diagram is recalculated. The grid is then regularised by moving 
each grid point slightly toward the centroid of the corresponding Voronoi cell several 
times in an iterative process. The result is a random grid consisting of approximately 
regular polyhedra (see Brinch \&
Hogerheijde~\cite{bh10} for details). Ray-tracing is done by integrating straight lines 
of sight (rays) through the grid simply by summing up the constant contribution between 
the two intersections between a ray and the Voronoi facets for each Voronoi cell the ray 
passes through.

For the dust polarisation calculations, the main extension to LIME is to
include Stokes $Q$ and $U$ in the source function. 
We followed the
procedure developed by Lee \& Draine (\cite{ld85}), then by Wardle \& 
K\"onigl (\cite{wk90}), Fiege \& Pudritz (\cite{fp00}), and Padoan et al. 
(\cite{pg01}). As in Gon\c calves et al. (\cite{gg05}) and Frau et al.
(\cite{fg11}), hereafter GGW05 and FGG11, respectively, 
our code accounts for the dependence of the emerging continuum radiation 
on dust temperature 
distribution ($T_{\rm dust}$)
supplied by the user (e.g., calculated self-consistently through tools such 
as RADMC/RADMC-3D\footnote{\tt www.ita.uni-heidelberg.de/$\thicksim$dullemond/software/radmc-3d}; Dullemond \& Dominik~\cite{dd04}).
According to this method, the polarisation degree $p$ and the polarisation 
(or position) angle 
$\chi$, that is the direction of polarisation in the plane of the sky, are given by

\be\label{Eqp}
p=\frac{\sqrt{q^{2}+u^{2}}}{i}
\ee
and
\be
\tan2\chi=\frac{u}{q}\,,
\ee
respectively. $q$, $u$, and $i$ are called ``reduced'' Stokes parameters and 
are defined in terms of the standard Stokes parameters
through a constant which depends on the cross section, a polarisation reduction 
factor due to imperfect grain alignment and
to the turbulent component of the magnetic field, and the Planck function. 
The ``reduced'' parameters are described by 

\be
q=\int_{0}^{\infty}\alpha_{\rm max}\ B_\lambda(T_{\rm dust})\ e^{-\tau_\lambda}\,\cos2\psi\,\cos^{2}\gamma\,\ud\tau_\lambda\,,
\ee
\be
u=\int_{0}^{\infty}\alpha_{\rm max}\ B_\lambda(T_{\rm dust})\ e^{-\tau_\lambda}\,\sin2\psi\,\cos^{2}\gamma\,\ud\tau_\lambda\,,
\ee
and
\be
i=\Sigma-\Sigma_{2}\,,
\ee
where $\psi$
is the angle between the north direction in the plane of the sky 
and the component of the magnetic field ({\bf B}) in this plane
and $\gamma$ is
the angle between 
the direction of {\bf B} and the plane of the sky.
$\Sigma$ and $\Sigma_{2}$ are given by

\be
\Sigma=\int _0 ^\infty B_\lambda(T_{\rm dust})\ e^{-\tau_\lambda}\,\ud\tau_\lambda=I_\lambda
\ee
and
\be\label{EqSigma2}
\Sigma_{2}=\frac{1}{2}\int_{0}^{\infty}\alpha_{\rm max}\ B_{\lambda}(T_{\rm dust})\ e^{-\tau_{\lambda}}\ \left(\cos^{2}\gamma-\frac{2}{3}\right)\,\ud\tau_{\lambda}\,,
\ee
where $\alpha_{\rm max}$ accounts
for grain cross sections and alignment properties.
For polarised 
grains, the effective cross section is given
by the difference between the extinction cross section and a contribution coming
from polarisation. For grains without polarisation properties, Stokes $I$ is
basically given by $\Sigma$.
Indeed, the
grain properties may vary with density
(Fiege \& Pudritz~\cite{fp00}), 
but in the absence of detailed information, unless otherwise specified,
we assume $\alpha_{\rm max}$ to be constant and equal to
0.15 in the remainder of this paper, corresponding to a maximum polarisation degree of about $15\%$, 
as suggested by GGW05. 
Higher values has not been observed and lower values appear unable to reproduce
observations of the magnetic field of prestellar cores 
(FGG11).
However, it is important to notice that the {\tt DustPol} module
allows the maximum polarisation degree to vary 
according to a user defined function (e.g. density or temperature dependent).

After the ray-tracing, the code creates a file in FITS format which consists of three slices, one 
for each Stokes parameters. This FITS file can be read by the 
{\tt sim\_observe-sim\_analyze} tasks in
the 
CASA package to prefigure the expected polarised flux intensity map obtained by ALMA.


\section{Applications}\label{applications}
In this Section, a few applications of the {\tt DustPol} routines are presented, showing their
capabilities to manage input from analytical models as well as numerical simulations.
A comparison with SMA available observations is also provided.

\subsection{Toy models}
It is instructive to start by considering geometrical models to check the effective
accuracy of the code. 
We examine sources with a 
purely toroidal ($\mathrm{\mathbf{B}_{tor}}$), a dipole ($\mathrm{\mathbf{B}_{dip}}$),
and a quadrupole ($\mathrm{\mathbf{B}_{quad}}$) magnetic field whose 
spherical components ($B_{r},B_{\theta},B_{\phi}$) are

\begin{equation}
\mathbf{B}_{\rm tor}\propto(0,0,r^{-1})\,,
\end{equation}

\begin{equation}
\mathbf{B}_{\rm dip}\propto
\left(-\frac{2\cos\theta}{r^3},-\frac{\sqrt{1-\cos^2\theta}}{r^3},0\right)\,,
\end{equation}
and

\begin{equation}
\mathbf{B}_{\rm quad}\propto
\left(-\frac{3(3\cos^2\theta-1)}{2r^4},
-\frac{3\cos\theta\sqrt{1-\cos^2\theta}}{r^4}
,0\right)\,,
\end{equation}
respectively.
%
%
We assumed a Bonnor-Ebert sphere (BES) density profile
(Bonnor~\cite{b56}; Ebert~\cite{e55}) with a central density of 
$2\times10^{13}$ m$^{-3}$, a radius of the inner flat region of 500~AU, and an
outer radius of 0.1~pc. We adopted
a wavelength of 850~$\mu$m, a 
constant dust temperature equal to 10~K for a source at the distance of 140~pc, and a
resolution of $0\farcs5$. Images have then been smoothed to a beam of
$10^{\prime\prime}$ to get sharper contours. 
The magnetic field configurations are shown for four
different inclinations with respect to the plane of the sky, from zero (face-on
source) to $\pi/2$ radians (edge-on source). 

\begin{figure}[!h] 
\resizebox{\hsize}{!}{\includegraphics[angle=0]{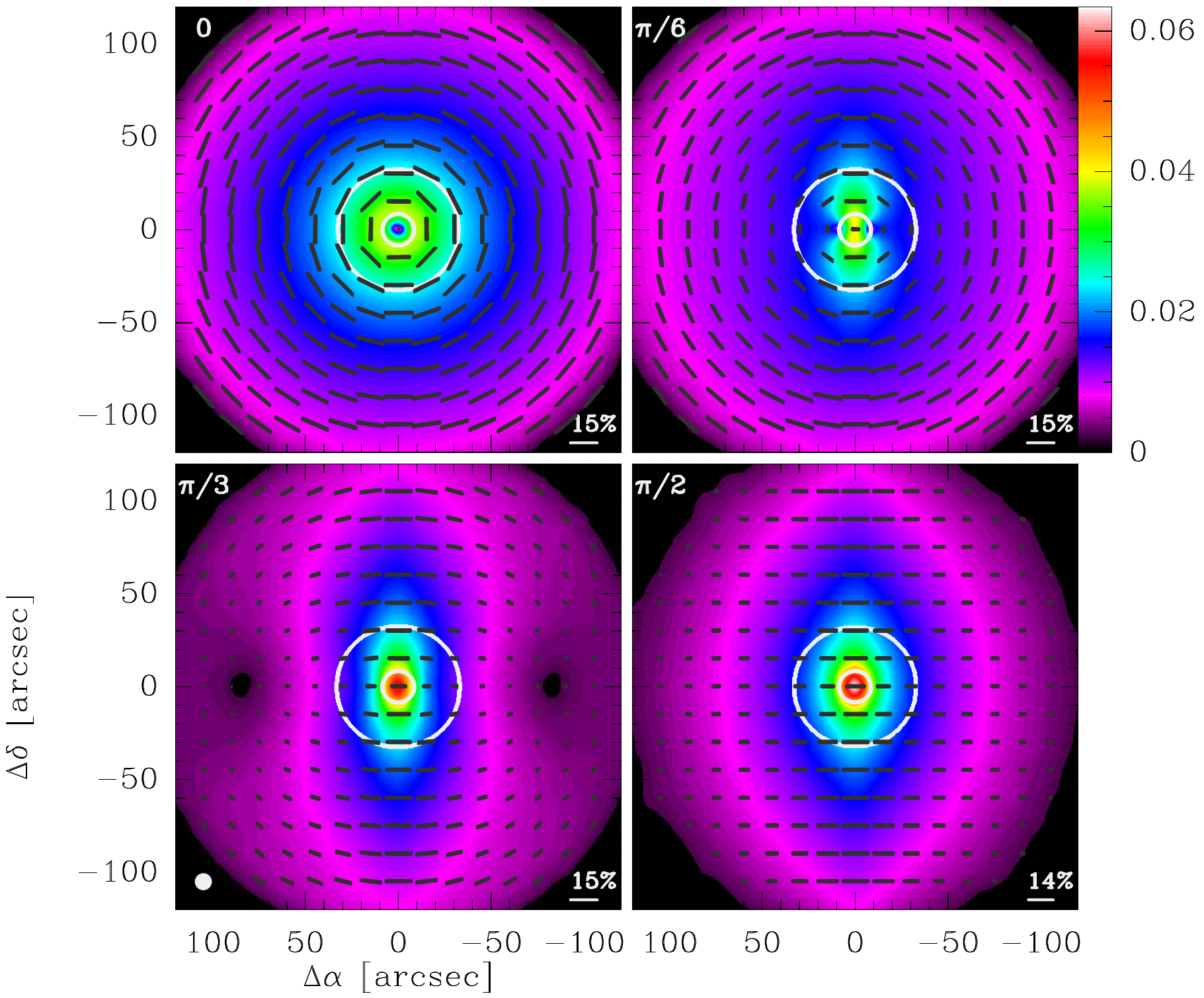}} 
\caption{Polarisation maps for a purely toroidal magnetic field and a BES density
distribution (model radius, $r_{\rm model}=2.25\times10^{4}$~AU). 
{\em Black vectors} represent the direction of the magnetic field, whose length is 
proportional to the fractional 
polarisation, obtained by flipping by 90 degrees the dust polarisation vectors.
{\em White contours} show 50  
and 90 per
cent
of the 850-$\mu$m dust emission (Stokes $I$) peak and the 
{\rm coloured map} depicts the polarised flux intensity ($I_{p}$). 
The scale bar to the right gives $I_{p}$ in 
Jy/(10$^{\prime\prime}$ beam). The inclination of the models with respect to the plane
of the sky is given in the upper left
corner of each subplot. The synthesised beam is shown in the lower left corner.
}
\label{19028fg1} 
\end{figure}

Figure~\ref{19028fg1} refers to a purely toroidal magnetic field. As expected,
for a face-on source field lines are concentric while,
approaching to the edge-on configuration, they become parallel in the plane of the
sky and can be pictured as concentric rings perpendicular to the vertical axis of
symmetry of the source. It can be easily understood that the depolarisation
in the outer parts is due to the bending of the field lines which become 
perpendicular to the line of sight.

\begin{figure}[!h] 
\resizebox{\hsize}{!}{\includegraphics[angle=0]{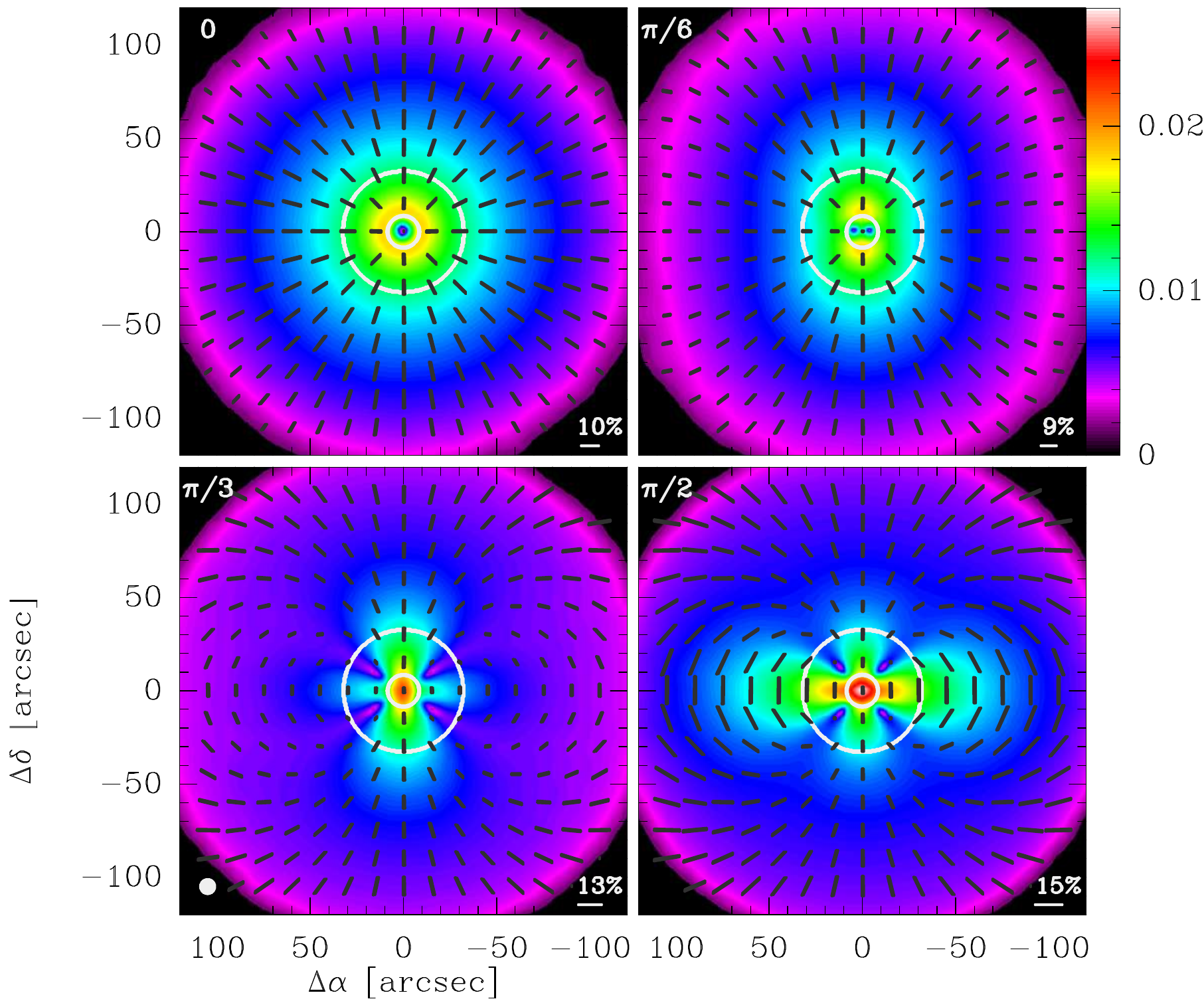}} 
\caption{Polarisation maps for a dipole magnetic field and a BES density
distribution ($r_{\rm model}=2.25\times10^{4}$~AU).
See Fig.~\ref{19028fg1} for further information.}
\label{19028fg2} 
\end{figure}

\begin{figure}[!h] 
\resizebox{\hsize}{!}{\includegraphics[angle=0]{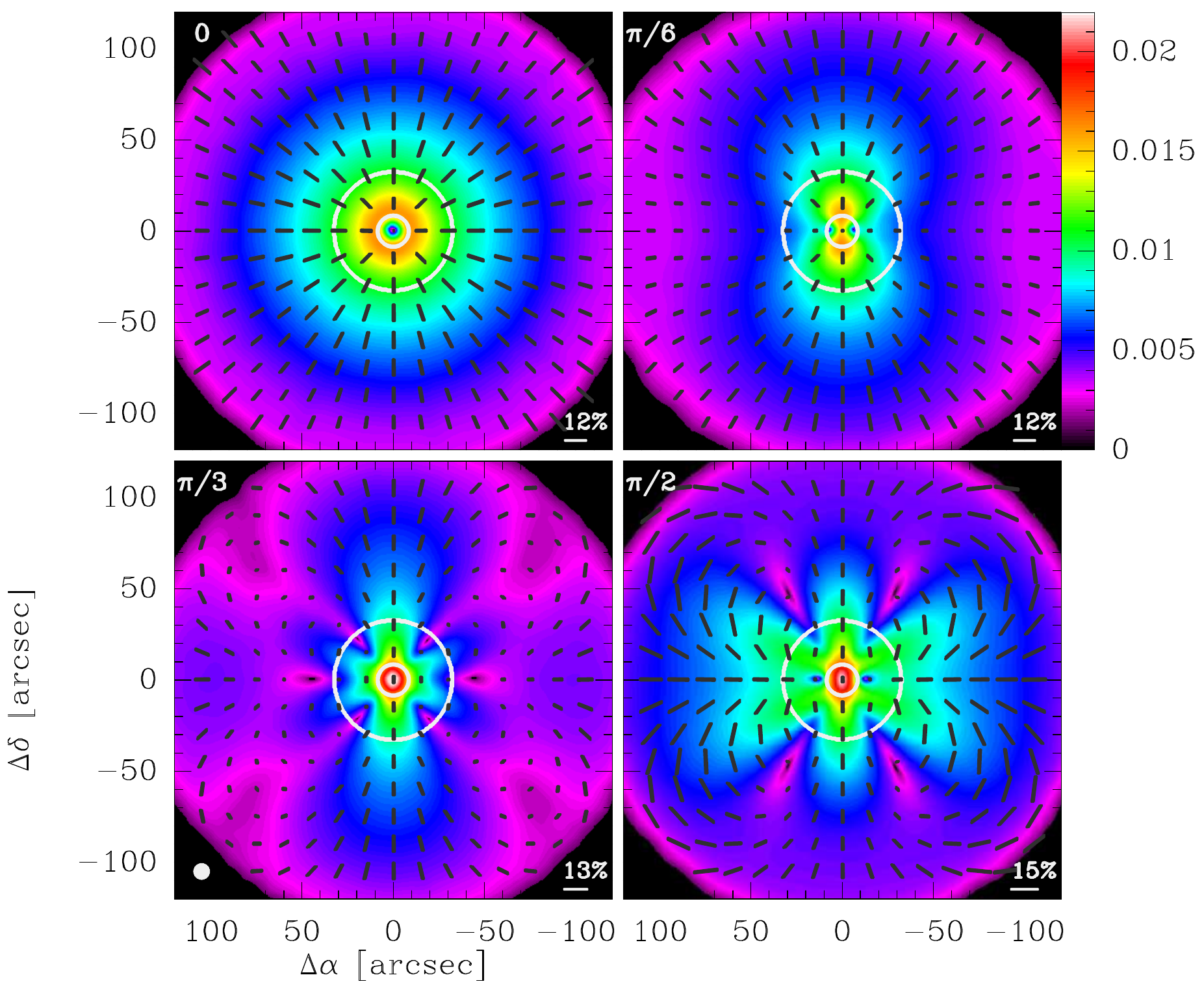}} 
\caption{Polarisation maps for a quadrupole magnetic field and BES density 
distribution ($r_{\rm model}=2.25\times10^{4}$~AU). 
See Fig.~\ref{19028fg1} for further information.}
\label{19028fg3} 
\end{figure}

Figures~\ref{19028fg2} and~\ref{19028fg3} show a radial configuration of
the polarisation vectors for a face-on source as well as the expected dipole and
quadrupole field shape arising as the inclination increases.

\subsection{Analytical models}\label{analyticalmodels}
\begin{figure}[!h] 
\resizebox{\hsize}{!}{\includegraphics[angle=0]{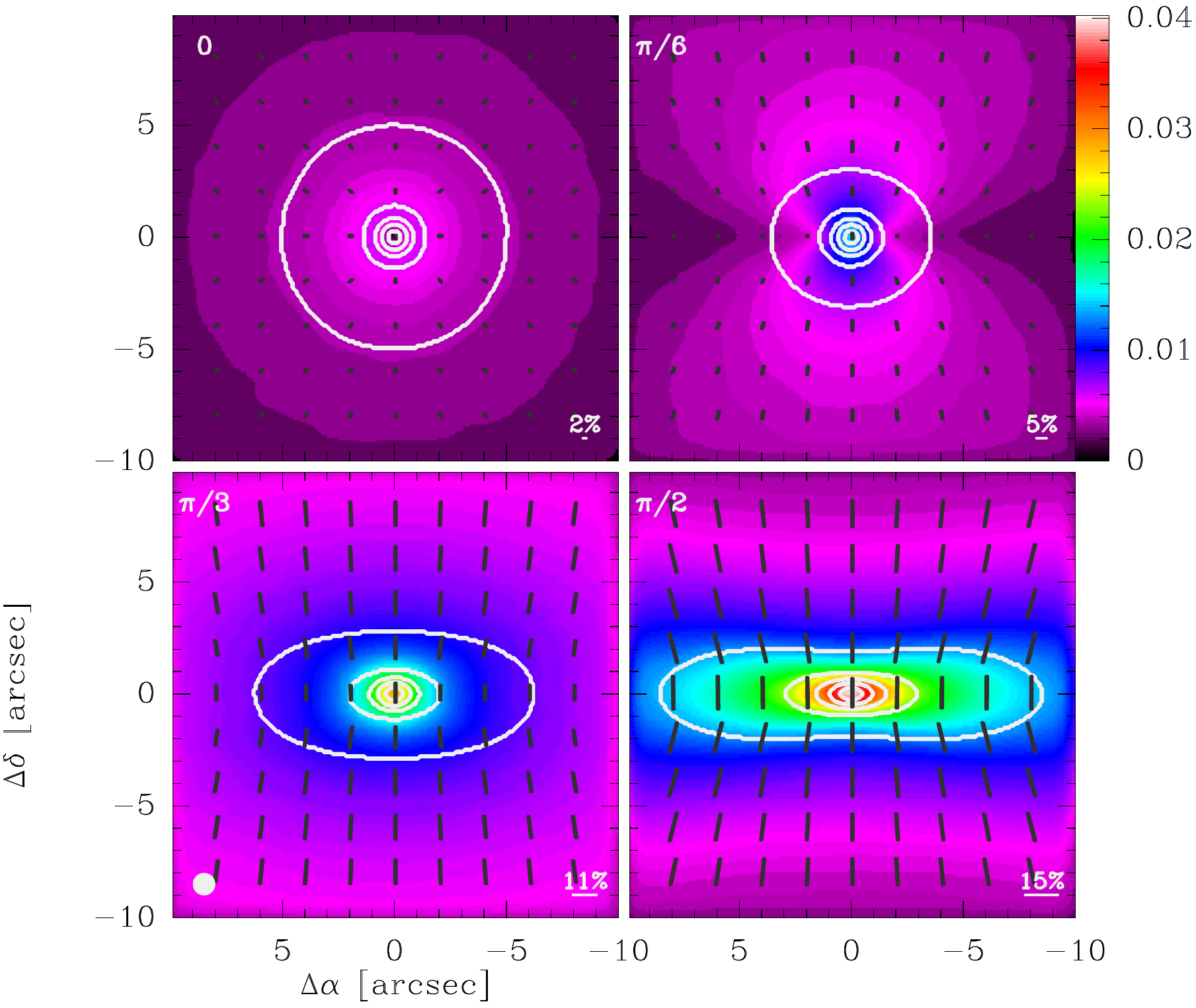}} 
\caption{Polarisation maps for the $H_0=1.25$ singular isothermal toroid according to 
the LS96 model
($r_{\rm model}=2.8\times10^{3}$~AU). 
{\em White contours} show 10, 30, 50, 70,  
and 90 per
cent of the 850-$\mu$m dust emission (Stokes $I$) peak and the 
scale bar to the right gives the polarised flux intensity in Jy/(1$^{\prime\prime}$
beam).
See Fig.~\ref{19028fg1} for further information.}
\label{19028fg4} 
\end{figure}

In order to test the correctness of our code, we reproduced the
model described in Li \& Shu (\cite{ls96}), hereafter LS96, 
to compare our results with those obtained 
by GGW05. 
This detailed comparison of the LS96 model between the 
results of GGW05 and our code is worthwhile to have a further validation of the
reliability of the code.
The model of LS96 is a semi-analytical, self-similar description of an axi-symmetric 
isothermal, self-gravitating protostellar core. It reproduces the hourglass magnetic field 
configuration suggested by
observations (Girart et al.~\cite{gr06}).
The governing equations hinge on a dimension-less parameter, $H_0$, 
which characterises the fraction of support provided against
self-gravity by poloidal magnetic fields relative to that of gas pressure
(see LS96 for more details). 
In particular, Fig.~\ref{19028fg4} shows the results for
a core with 
$H_0=1.25$, observed at 850~$\mu$m. 
We adopted a distance of 140~pc, $T_{\rm dust}=10$~K and an angular resolution of
$0\farcs1$. Images have been smoothed to $1^{\prime\prime}$ for plotting.

\begin{figure}[!h] 
\resizebox{.9\hsize}{!}{\includegraphics[]{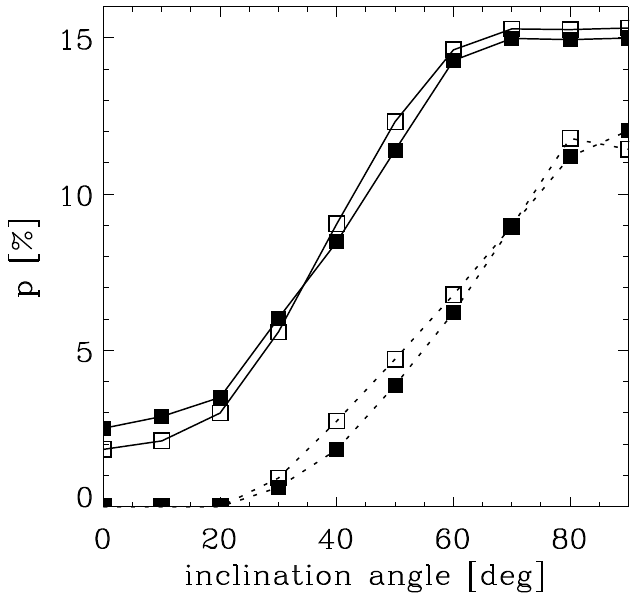}} 
\caption{Maximum ({\em solid lines}) and mininum ({\em dotted lines})
percentage of polarisation degree as a function of the inclination of the core
with respect to the plane of the sky.
{\tt DustPol} model, {\em solid squares}; GGW05 model,
{\em empty squares}.}
\label{19028fg5} 
\end{figure}

Our computation 
supports the conclusions of GGW05 concerning 
the depolarisation at intermediate inclinations. 
In the central regions
the bending of the field lines is stronger and the integration along line of sights
close to the centre leads to a lower polarisation degree. 
We compared the minimum and the maximum degree of polarisation as a function
of the
inclination of the core with values computed by GGW05.
For a direct comparison, we smoothed our model results to $12^{\prime\prime}$
which is the beam used in GGW05 and we obtain a substantial 
agreement as shown
in Fig.~\ref{19028fg5}.

\begin{figure}[!h] 
\resizebox{\hsize}{!}{\includegraphics[]{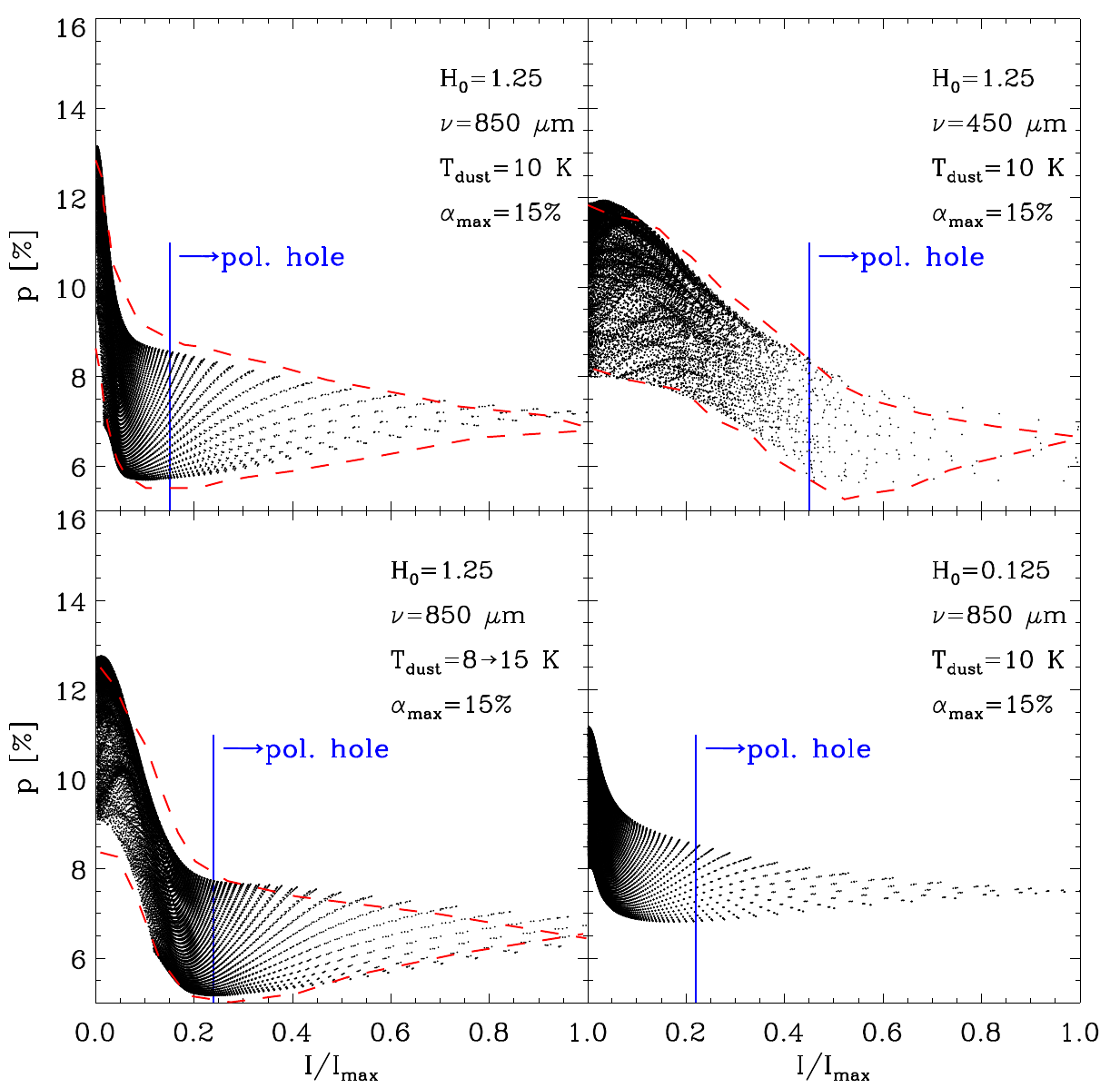}} 
\caption{Degree of polarisation ($p$) as a function of the intensity normalised to its
peak value ($I/I_{\rm max}$) for the LS96 toroid with an
inclination of 50 degrees with respect to the plane of the sky
for different wavelengths, temperatures, and 
values of $H_0$.
Each point in these diagrams represents a grid point from the model and
multiple tracks are an artefact due to the grid sampling.
The {\em red dashed lines} show the maximum and minimum values of the 
polarisation degree found by GGW05, see their Fig.~8.
Regions on the right of the {\em vertical blue solid lines} correspond to the extent of the
polarisation holes.}
\label{19028fg6} 
\end{figure}

We examined the dependence of the polarisation degree on wavelength, temperature, 
and
$H_0$, finding analogies with GGW05.
The two top panels of Fig.~\ref{19028fg6} show the dependence of the 
polarisation degree on the wavelength. 
Due to the fact that the emission for lower wavelengths is rather flat and less concentrated
(Gon\c calves et al.~\cite{gg04}), 
the region with low and uniform
values of $p$, the so-called polarisation hole, has a smaller extent in this Figure at 
450~$\mu$m with respect to 
850~$\mu$m. 
It follows that, if the polarisation hole has a
fixed physical radius, the corresponding value of $I/I_{\rm max}$ will be larger
at 450~$\mu$m than at 850~$\mu$m.
We also probed the dependence of the polarisation degree on the
dust temperature as shown by the comparison of the 
two left panels of Fig.~\ref{19028fg6}. 
The lower left panel 
shows the result for the assumption of an external heating with 
$T_{\rm dust}$ going from $\sim$15 to $\sim$8~K inside a radius of 0.1~pc
(Gon\c calves et al.~\cite{gg04}). As in GGW05, we conclude that a non-isothermal
distribution of the dust temperature may help the observation of the depolarisation
effect. 
In fact, the magnetic field component in the plane of the sky is generally larger far from the central thermal dust 
emission peak. Thus, when in presence of an interstellar radiation field, the contribution 
of the external layers of the prestellar core to 
the Stokes parameters becomes substantial.
Finally, we examined the dependence of the polarisation degree 
on $H_0$, which is altering the magnetic
field support. The lower right panel of Fig.~\ref{19028fg6} shows the distribution
of the polarisation degree for $H_0=0.125$, corresponding to a mild pinching of the
magnetic field lines. As found in GGW05, it is interesting to notice that even
a low value
of $H_0$ leads to modifications of the polarisation degree for increasing intensities,
measured as the difference between the maximum ($p_{\rm max}$) and the minimum ($p_{\rm min}$)
degree
of polarisation observed. For $H_0=0.125$, we found
$\Delta p=p_{\rm max}-p_{\rm min}=4.5\%$ while, for comparison, $\Delta p=7.6\%$ for 
$H_0=1.25$, that is for a more pinched field, assuming in both cases $\nu=850~\mu$m,
$T_{\rm dust}=10$~K, and $\alpha_{\rm max}=15\%$.

\begin{figure}[!h] 
\resizebox{\hsize}{!}{\includegraphics[]{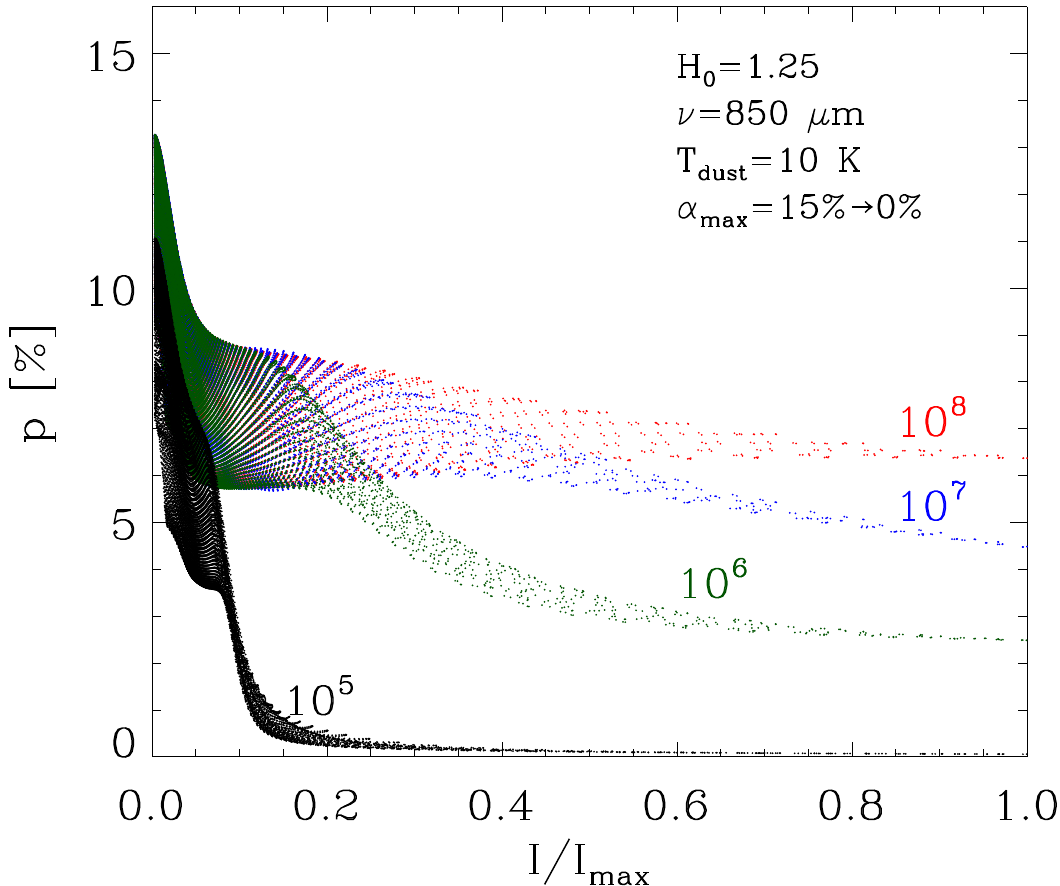}} 
\caption{Degree of polarisation ($p$) as a function of the intensity normalised to its
peak value ($I/I_{\rm max}$) for the LS96 toroid as in Fig.~\ref{19028fg6}.
This plot shows the decrease in $p$ for high values of
$I/I_{\rm max}$ due to the assumption of a step
function for the maximum degree of polarisation.
Labels ($10^{5}$ to $10^{8}$~cm$^{-3}$) represent the limit on $n$(H$_{2}$) above which
$\alpha_{\rm max}=0$. 
}
\label{19028fg7} 
\end{figure}

The magnetic field geometry and temperature distribution are important components for the description of the observed values of the polarisation degree, but they cannot account for 
the whole set of observational constraints. 
For example, it is known that the maximum degree of 
polarisation, $\alpha_{\rm max}$, is strongly dependent on the cutoff in the grain size 
distribution (Cho \& Lazarian~\cite{cl05}) and depolarisation may be the result of 
alterations in the grain structure at higher densities which may reduce the dust grain 
alignment efficiency (Lazarian \& Hoang~\cite{lh07}).
In order to 
test the dependence of the polarisation on grain properties,
we assumed a step function for 
$\alpha_{\rm max}$,  
which drops from 15\% to zero for molecular hydrogen number densities higher 
than 10$^5$, 10$^6$, 10$^7$, and 10$^8$~cm$^{-3}$. Results are shown in
Fig.~\ref{19028fg7}. As expected, the polarisation degree 
decreases with intensity more rapidly as the limit on $n$(H$_{2}$) decreases. Despite 
the assumption of the step function is purely arbitrary, it is possible to deduce that 
the case of no polarisation for 
$n$(H$_{2})>10^5$~cm$^{-3}$ 
has to be discarded since there is no observational evidence that $p$ is zero for $I\gtrsim0.2I_{\rm max}$. 
Although the theory of grain alignment is still not completely developed, the extreme
flexibility in the parameterisation of $\alpha_{\rm max}$ allows the user to easily
explore the parameter space.

\subsection{Numerical models}\label{numericalmodels}
The novelty of {\tt DustPol} is that it can easily manage inputs from numerical 
simulations set out on regular or irregular grids. In this paper we give a couple of
examples of 3D numerical calculations of a collapsing magnetised dense core
(Hennebelle \& Fromang~\cite{hf08}, Hennebelle \& Ciardi~\cite{hc09}) 
which make use of the adaptive
mesh refinement (AMR) code RAMSES (Teyssier~\cite{t02}, Fromang et al.~\cite{fh06}).
We describe only qualitatively the results postponing quantitative 
conclusions for a following paper (Padovani et al. {\em in prep.}).
 
The grid of a numerical simulation 
is sampled as if it were an analytic function. 
Each point of the grid built by LIME takes the value of density, temperature, and
magnetic field of the closest model point. Before this step, 
the output of a numerical simulation has to be converted in the format
readable by LIME. To this end, a
simple wrapper is distributed together with {\tt DustPol}.

\begin{figure}[!h] 
\resizebox{\hsize}{!}{\includegraphics[angle=0]{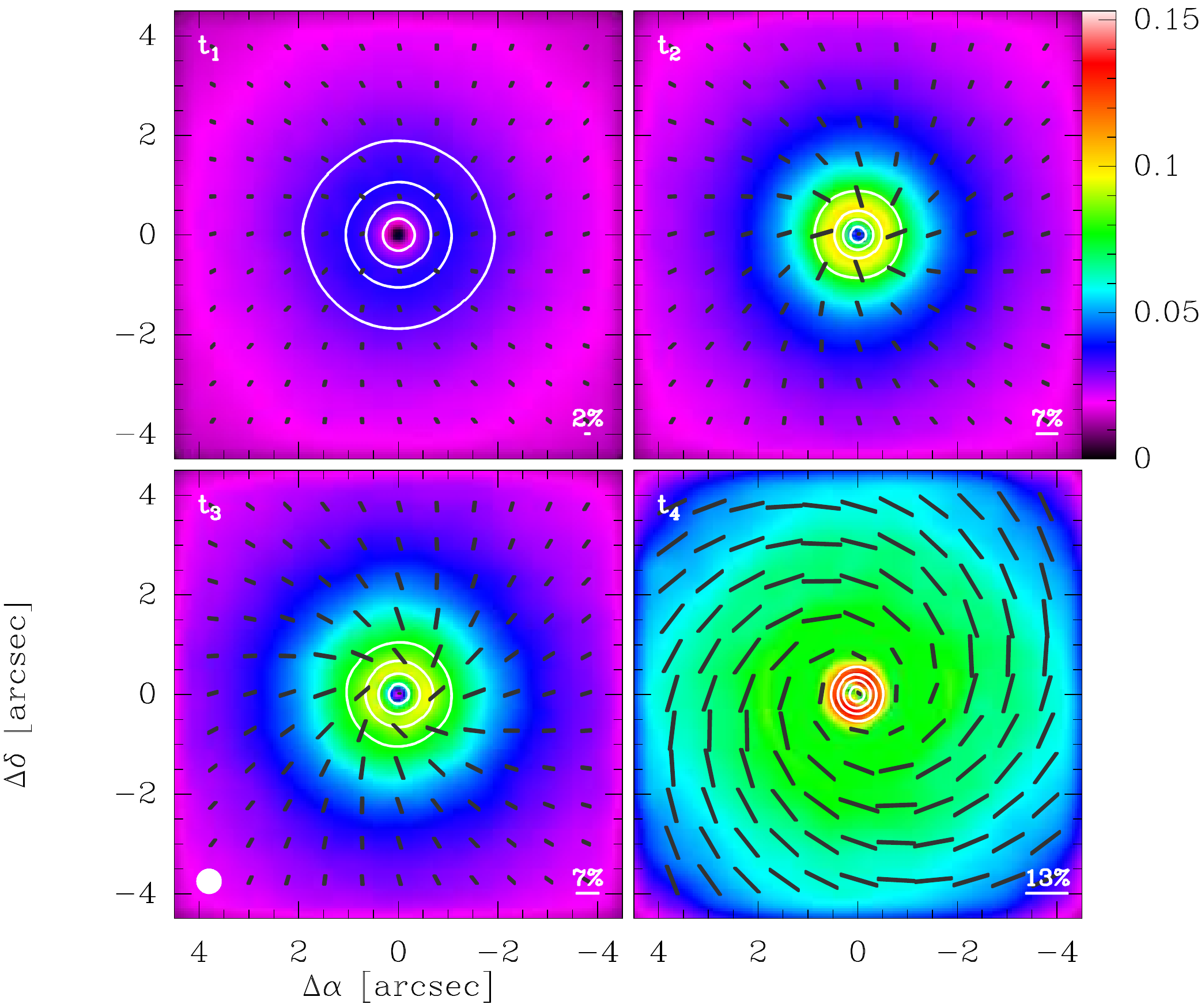}} 
\caption{Polarisation maps for a collapsing magnetised dense core 
(face-on view)
obtained with the
RAMSES code ($r_{\rm model}=6.3\times10^{2}$~AU). 
Labels refer to different times of the collapse:
$t_{1}=1.76\times10^{4}$~yr,   
$t_{2}=1.86\times10^{4}$~yr,  
$t_{3}=1.94\times10^{4}$~yr, and  
$t_{4}=2.96\times10^{4}$~yr.
{\em White contours} show 30, 50, 70,  
and 90 per
cent of the 850-$\mu$m dust emission (Stokes $I$) peak and
the scale bar to the right gives $I_{p}$ in Jy/($0\farcs5$~beam).
See Fig.~\ref{19028fg1} for further information.}
\label{19028fg8} 
\end{figure}

RAMSES was run for a
collapsing rotating core with mass-to-flux ratio equal to 5. The magnetic field and
the angular momentum vectors are initially aligned. We extracted four snapshots from
the RAMSES run at four times (see caption of Fig.~\ref{19028fg8}), 
chosen
in order to predict the evolution of the polarisation since magnetic field
lines get tangled as the core rotates.
The output of the RAMSES run at these time samples was used as input to
{\tt DustPol} using
a distance of 140~pc, 
with a wavelength of 850~$\mu$m, $T_{\rm dust}=10$~K, and an angular resolution
of $0\farcs05$, then smoothed at $0\farcs5$, which is a 
resolution typically achievable by ALMA. 
Note that these maps are not corrected for the instrumental response of the
interferometer. 
Initially, the magnetic field has hourglass shape, as time increases, the field
configuration becomes purely toroidal.
Figure~\ref{19028fg8} shows the time evolution for a face-on
configuration, which yields an initially radial pattern changing to a more toroidal 
structure.

\begin{figure}[!h] 
\resizebox{\hsize}{!}{\includegraphics[angle=0]{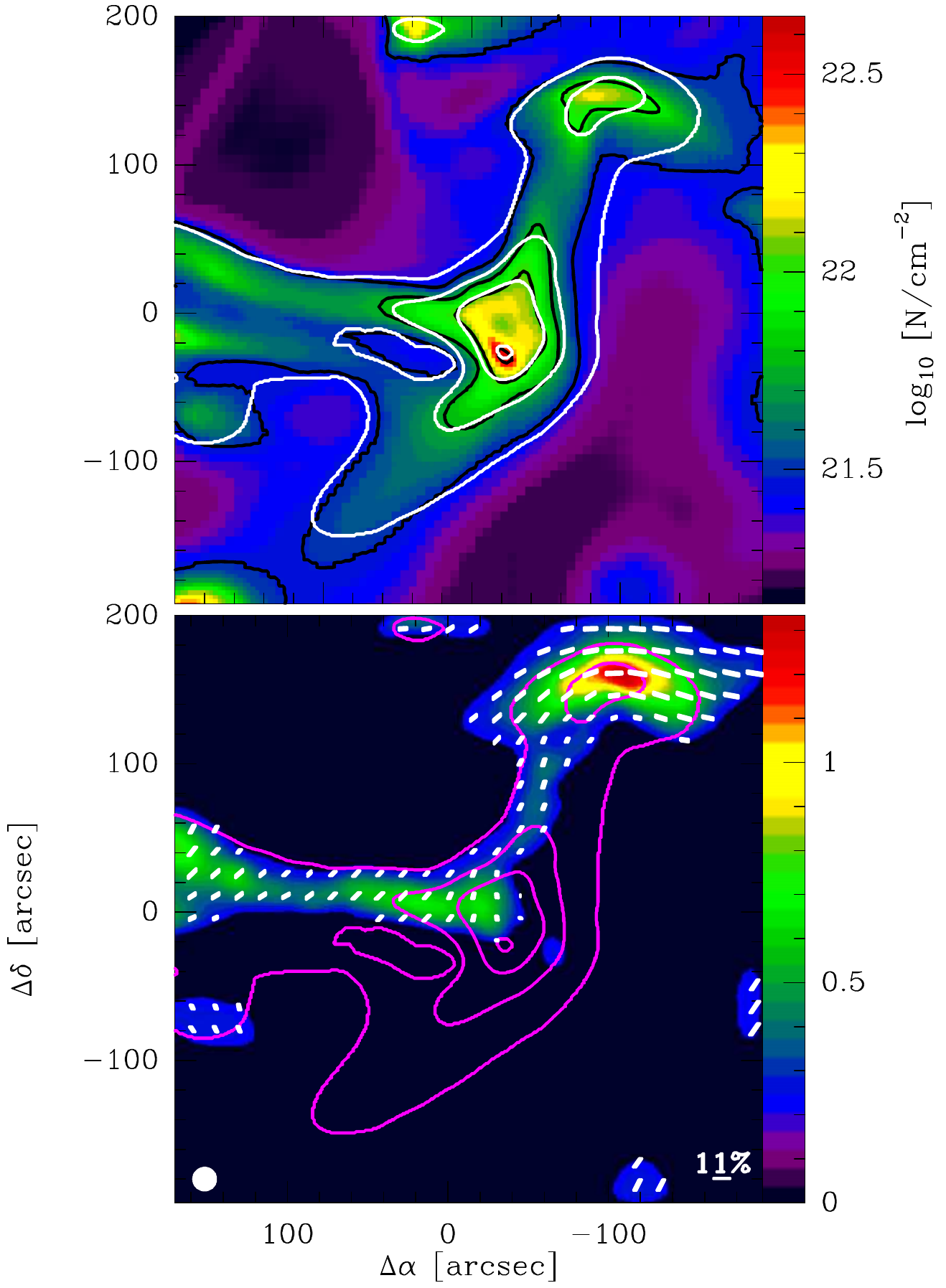}} 
\caption{{\em Upper panel}: column density distribution of molecular hydrogen for a 
prestellar core 
extracted from the StarFormat database.
{\em Black} and {\em white contours} 
show 30, 50, 70, and 95 per cent of the H$_{2}$ column density peak and of the
850-$\mu$m dust emission (Stokes $I$) peak, respectively.
The scale bar to the right gives $N($H$_{2}$) in logarithmic scale. {\em Lower panel}: 
polarisation map for the core depicted in the upper panel 
($r_{\rm model}=2.8\times10^{4}$~AU).
The scale bar to the right gives $I_{p}$ in Jy/($14\farcs5$~beam) and it
refers to the coloured map which has been debiased for values
below $3\sigma$.
{\em Magenta contours} show 30, 50, 70, and 95 per cent
of the 850-$\mu$m dust emission (Stokes $I$) peak; {\em white vectors}
show the
direction of the magnetic field.
The synthesised beam is shown in the lower left corner.}
\label{19028fg9} 
\end{figure}

Finally, we tested {\tt DustPol} using a 
large scale RAMSES
simulation 
(Dib et al.~\cite{dh10}) 
relative to a prestellar core with a mass of
1.62 M$_{\odot}$,
which has been obtained from
the ``Molecular cloud evolution with decaying turbulence'' project of 
the StarFormat database\footnote{\tt http://starformat.obspm.fr/starformat/projects}. 
The upper panel of Fig.~\ref{19028fg9} shows the column density 
distribution of molecular
hydrogen, $N($H$_2$), of this core.
{\tt DustPol} has been used to get the polarisation pattern using a
distance of 140~pc, $T_{\rm dust}=10$~K, 
and a wavelength of 850~$\mu$m. The
resulting image is shown in the lower panel of Fig.~\ref{19028fg9}.
It has been smoothed
from 1$^{\prime\prime}$ to $14\farcs5$ which is the beam
of SCUBA, accounting for the error beams 
for an accurate description of the beam profile (Hogerheijde \& Sandell~\cite{hs00}).

Under the optically thin assumption, the agreement between the RAMSES 
H$_{2}$ column density distribution and the {\tt DustPol} Stokes $I$ map is 
remarkable. In fact, 
after normalising $N($H$_2$) and Stokes $I$ with respect to their maximum values, 
we computed the ratio between $N($H$_2$) and Stokes $I$ finding 
a mean value of $0.95\pm0.12$. 
Additionally, it is worth noting the displacement between the
polarised emission and the dust continuum emission near the offset $(0,0)$ and,
conversely, the good overlap near $(-100,150)$. This can be easily understood by
looking at the local direction of the magnetic field into the RAMSES data cube.
We found that the magnetic field points towards the line
of sight around the centre of the map, thus 
explaining the decrease in polarisation degree. On the other hand,
the direction of the
magnetic field mainly lays on the plane of the sky near
the north-west clump, where the polarisation is stronger.

Working in tandem, {\tt DustPol} and StarFormat (to be publicly available) can be used
to interpret large-scale polarisation maps of molecular clouds, investigating
on magnetic field morphologies.

\subsection{Comparison with observations: NGC~1333 IRAS~4A}\label{compobs}
The ultimate 
goal of the {\tt DustPol} project is to generate simulations capable of being compared 
to the real data. Its main advantage is that it provides quick convergence because of
the adaptive gridding and it easily accommodates outputs from other codes as 
shown in Sect.~\ref{analyticalmodels} and~\ref{numericalmodels}.
Recently, FGG11 have developed a method to compare the observational 
polarisation maps of the low-mass Class~0 object NGC~1333 IRAS~4A 
to a set of magnetohydrodynamic (MHD) simulations, which allows
to find the best set of models that fit the magnetic field properties of NGC~1333 IRAS~4A.
They generate a synthetic source based on models and simulations and,
using the same equation set described in this work,
they provide synthetic Stokes maps.
These synthetic maps are convolved by the interferometric response in order to 
obtain the spatially filtered maps comparable to the real data.
To assess the quality of the {\tt DustPol} module, we repeated this scheme 
replacing the radiative transfer tool in FGG11 with {\tt DustPol}.
We used the Allen et al.~(\cite{as03}) model datacube with the best 
fitting parameters found by FGG11, namely
a mass of 1.6~M$_\odot$, a mean volume density within the angular size of 100~AU 
of $3\times10^{8}$~cm$^{-3}$, an inclination of 45 degrees with respect to
the plane of the sky, $H_0=0.125$ (see Sect.~\ref{analyticalmodels}),
and the dust temperature profile derived by Maret et al.~(\cite{mc02}). 
We ran {\tt DustPol} with a resolution of 
$0\farcs16$, assuming a source 
distance of 300~pc, and a wavelength of 880~$\mu$m.
Then, we applied the same convolution tool developed first by Gon\c calves 
et al. (\cite{gg08})
and improved by FGG11 to the output map of {\tt DustPol} for
a final synthesised beam of $1\farcs24\times1\farcs12$.
Figure~10 
shows our results, to be compared with the third row of 
Fig.~8 in FGG11.
The observations and the model 
are in good agreement and the general morphologies of the total emission
(Stokes $I$) and the polarised emission are consistent. 
In fact, even if the polarisation vectors in the map generated by {\tt DustPol} cover a 
larger area, the difference map between the Stokes $Q$ (and $U$) maps created by the 
two radiative transfer tools show no features above the 3$\sigma$ level
($\sigma=2.5$~mJy beam$^{-1}$). 
Therefore, the differences are due to the statistical noise at the low RMS regions.
Finally, the differences between the position angles evaluated with the radiative
transfer tool in FGG11 and {\tt DustPol} have a standard deviation of 0.48 degrees,
confirming the reliability of {\tt DustPol}.

\begin{figure}[!h]\label{19028fg10}  
\resizebox{0.9\hsize}{!}{\includegraphics[angle=0]{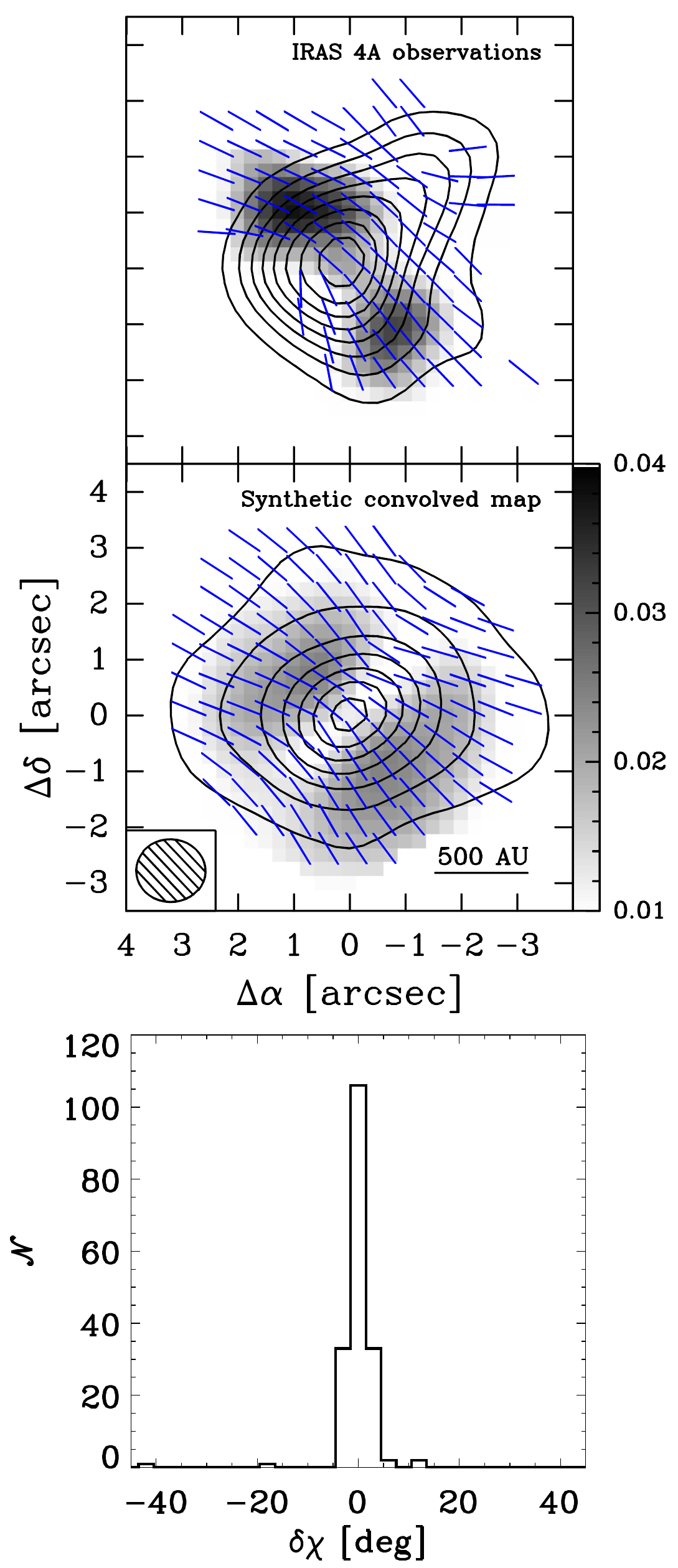}}
\caption{
{\em Top panel}: dust polarisation SMA maps toward NGC~1333 IRAS~4A from FGG11. 
{\em Middle panel}: synthetic dust polarisation map generated with {\tt DustPol}, 
following the FGG11 convolution scheme, using the best fitting parameters of the 
Allen et al.~(\cite{as03}) model found by 
FGG11 (see Sect.~\ref{compobs} for more details).
Each panel shows the intensity ({\em contours}), polarised intensity 
({\em pixel map}), and direction of magnetic field ({\em blue segments}).
Contours depict emission levels from 6$\sigma$ up to the maximum value in steps of 
6$\sigma$, where $\sigma = 0.02$~Jy~beam$^{-1}$. The middle panel shows the beam and 
the angular and spatial scale.
{\em Bottom panel}: histogram of the difference between the position angles computed
with the radiative transfer tool in FGG11 and {\tt DustPol}.}
\end{figure} 

\section{Conclusions}\label{conclusions}
We have presented the {\tt DustPol} module of the
Adaptable Radiative Transfer Innovations for Submillimetre Telescopes (ARTIST)
package, a tool for
modelling the polarisation of thermal dust emission
self-consistently. The module is an extension to the
radiative transfer code LIME making it possible to 
model
complex three-dimensional structures.
The synergy between {\tt DustPol} and codes such as RAMSES and RADMC-3D will help
to enlighten the interpretation of the role of magnetic fields in star-forming
regions.

{\tt DustPol} can easily handle analytical as well as pre-gridded models
and outputs from other codes, accounting for
the dependence on dust temperature and, for the first time to our knowledge,
for variations of the polarisation degree.
A number of examples (Sect.~\ref{applications}) demonstrate 
the capabilities and the accuracy of {\tt DustPol} to interpret existing submillimetre 
data as well as new data from ALMA and the Planck satellite.
The output of {\tt DustPol} is a FITS file whose header is consistent with the format
of the input models for the 
{\tt sim\_observe-sim\_analyze} tasks of the CASA package. This makes it 
straightforward, e.g., to make prediction about the polarised flux for a
given configuration of the ALMA antennas.

The code is publicly available as long as appropriate reference is
made to this paper. It is accessible through the
complete ARTIST package, taking advantage of the Python-based graphic user interface
that includes a model library or through a stand-alone version.

\acknowledgements
MP would like to acknowledge Daniele Galli for many 
helpful and inspiring discussions on development of the code and on drafting the paper.
We also want to thank the referee, Karl Misselt, for careful reading of the manuscript 
and helpful comments.
This work was funded as part of the ASTRONET 
initiative by the Spanish (MINECO), 
Dutch (NWO), and German (BMBF) funding agencies.
MP, JMG and PF are supported by MINECO grants AYA2008-04451-E, 
AYA2008-06189-C03-02, and AYA2011-30228-C03-02 (Spain) and by AGAUR grant 2009SGR1172 (Catalonia). 
PF is partially supported by MINECO fellowship 
FPU (Spain). 
The research in Copenhagen is supported by a Junior Group Leader Fellowship from the 
Lundbeck Foundation to JKJ as well as by the Danish National Research Foundation through 
the establishment of Centre for Star and Planet Formation.
Part of the time, RK is financially supported by grant "LPDS 2011-5" of the 
Leopoldina Fellowship Programme. 
WV acknowledges support by the Deutsche 
Forschungsgemeinschaft (DFG; through the Emmy Noether Research grant VL 61/3-1).


\begin{thebibliography}{}


\bibitem[2003]{as03}
Allen, A., Shu, F.~H. \& Li, Z.
2003, \apj, 599, 351 

\bibitem[1997]{b97}
Basu, S.
1997, \apj, 485, 240

\bibitem[1956]{b56}
Bonnor, W.~B.,
1956, MNRAS, 116, 351

\bibitem[2010]{bk10}
Boss, A.~P., Keiser, S.~A., Ipatov, S.~I., Myhill, E.~A. \& Vanhala, H.~A.~T.
2010, \apj, 708, 1268

\bibitem[2010]{bh10}
Brinch, C. \& Hogerheijde, M.
2010, \aap, 523, 25

\bibitem[2005]{cl05}
Cho, J. \& Lazarian, A.
2005, \apj, 631, 361

\bibitem[1996]{ct96}
Crutcher, R.M., Troland, T.H., Lazareff, B. \& Kaz\`es, I. 
1996, \apj, 456, 217

%
\bibitem[2011]{dn11}
Davidson, J.~A., Novak, G., Matthews, T.~G., Matthews, B., Goldsmith, P.~F.,
Chapman, N., Volgenau, N.~H., Vaillancourt, J.~E. \& Attard, M.
2011, \apj, 732, 97 

\bibitem[1951]{dg51}
Davis, L. \& Greenstein, J.~L.
1951, \apj, 114, 206

\bibitem[2010]{dh10}
Dib, S., Hennebelle, P., Pineda, J.~E., Csengeri, T., Bontemps, S., 
Audit, E. \& Goodman, A.~A.
2010, \apj, 723, 425

\bibitem[1976]{dm76}
Dolginov, A.~Z., \& Mytrophanov, I.~G.
1976, \apss, 43, 257 

\bibitem[1996]{dw96}
Draine, B.~T. \& Weingartner, J.~C.
1996, \apj, 470, 551

\bibitem[2004]{dd04}
Dullemond, C.~P. \& Dominik, C. 
2004, \aap, 417, 159

\bibitem[1955]{e55}
Ebert, R.,
1955, Z. f\"ur Astr., 37, 217

\bibitem[2002]{e02}
Efroimsky, M.
2002, \apj, 575, 886

\bibitem[1993]{e93}
Elmegreen, B.~G.
1993, \apj, 419, L29

\bibitem[2008]{ft08}
Falgarone, E., Troland, T.H., Crutcher, R.M. \& Paubert, G. 
2008, \aap, 487, 247

\bibitem[2000]{fp00}
Fiege, J.~D. \& Pudritz, R.~E. 
2000, \apj, 544, 830

\bibitem[1998]{fl98}
Fogel, M.~E. \& Leung, C.~M.
1998, \apj, 501, 175

\bibitem[2011]{fg11}
Frau, P., Galli, D. \& Girart, J.~M.
2011, \aap, 535, 44 (FGG11)

\bibitem[2006]{fh06}
Fromang, S., Hennebelle, P. \& Teyssier, R.
2006, \aap, 457, 371

\bibitem[1996]{go96}
Gammie, C.~F. \& Ostriker, E.~C.
1996, \apj, 466, 814

\bibitem[1999]{gc99}
Girart, J.~M., Crutcher, R.~M., \& Rao, R.
1999, \apjl, 525, L109 

\bibitem[2006]{gr06}
Girart, J.~M., Rao, R. \& Marrone, D.~P. 
2006, Science, 313, 812

\bibitem[2009]{gb09}
Girart, J.~M., Beltr\'an, N., Zhang, Q., Rao, R. \& Estalella, R. 
2009, Science, 324, 1408

\bibitem[1951]{g51}
Gold, T.
1951, Nature, 169, 322

\bibitem[2004]{gg04}
Gon\c calves, J., Galli, D. \& Walmsley, C.~M. 
2004, \aap, 415, 617

\bibitem[2005]{gg05}
Gon\c calves, J., Galli, D. \& Walmsley, C.~M. 
2005, \aap, 430, 979 (GGW05)

\bibitem[2008]{gg08}
Gon\c calves, J., Galli, D. \& Girart, J.~M. 
2008, \aap, 490, L39

\bibitem[2008]{hf08}
Hennebelle, P. \& Fromang, S.
2008, \aap, 477, 9

\bibitem[2009]{hc09}
Hennebelle, P. \& Ciardi, A.
2009, \aap, 506L, 29

\bibitem[2000]{hs00}
Hogerheijde, M. \& Sandell, G.
2000, \apj, 534, 880


\bibitem[2003]{lg03}
Lai, S.-P., Girart, J.~M. \& Crutcher, R.~M.
2003, \apj, 598, 392 

\bibitem[2003]{l03}
Lazarian, A.
2003, Journal of Quantitative Spectroscopy \& Radiative Transfer, 79, 881

\bibitem[2007]{l07}
Lazarian, A.
2007, Journal of Quantitative Spectroscopy \& Radiative Transfer, 106, 225

\bibitem[2007]{lh07}
Lazarian, A. \& Hoang, T. 
2007, \mnras, 378, 910

\bibitem[1985]{ld85}
Lee, H.~M. \& Draine, B.~T.
1985, \apj, 290, 211

\bibitem[1996]{ls96}
Li, Z.-Y. \& Shu, F.~H. 
1996, \apj, 472, 211 (LS96)

\bibitem[2004]{mk04}
Mac Low, M.-M. \& Klessen, R.~S.
2004, RvMP, 76, 125

\bibitem[2002]{mc02}
Maret, S., Ceccarelli, C., Caux, E., Tielens, A.~G.~G.~M. \& Castets, A. 
2002, \aap, 395, 573

\bibitem[1971]{m71}
Martin, P.~G.
1971, \mnras, 153, 279

\bibitem[2001]{mw01}
Matthews, B.~C., Wilson, C.~D. \& Fiege, J.~D.
2001, \apj, 562, 400 

\bibitem[2002]{mw02}
Matthews, B.~C. \& Wilson, C.~D. 
2002, \apj, 574, 822

\bibitem[1956]{ms56}
Mestel, L. \& Spitzer, L.
1956, \mnras, 116, 503

\bibitem[2006]{mt06}
Mouschovias, T.~C., Tassis, K., \& Kunz, M.~W.
2006, \apj, 646, 1043
 
\bibitem[1989]{ng89}
Novak, G., Gonatas, D.~P., Hildebrand, R.~H., Platt, S.~R. \& Dragovan, M.
1989, \apj, 345, 802 

\bibitem[2001]{pg01}
Padoan, P., Goodman, A., Draine, B.~T., Juvela, M., Nordlund, \AA\ 
\& R\"ognvaldsson, \"O.~E. 
2001, \apj, 559, 1005


\bibitem[2008]{pg08}
Pinto, C. \& Galli, D.
2008, \aap, 484, 17

\bibitem[1979]{p79}
Purcell, E.~M.
1979, \apj, 231, 404

\bibitem[2009]{rg09}
Rao, R., Girart, J.~M., Marrone, D.~P., Lai, S.-P. \& Schnee, S.
2009, \apj, 707, 921 

\bibitem[2006]{ri06}
Ritzerveld, J. \& Icke, V. 
2006, Phys. Rev. E, 74, 26704

\bibitem[2011a]{sv11a}
Surcis, G., Vlemmings, W.~H.~T., Curiel, S., 
Hutawarakorn Kramer, B., Torrelles, J.~M. \& Sarma,  A.~P.
2011, \aap, 527, 48

\bibitem[2011b]{sv11b}
Surcis, G., Vlemmings, W.~H.~T., Torres, R.~M., van Langevelde, H.~J. \& Hutawarakorn Kramer, B.
2011, \aap, 533, 47

\bibitem[2009]{th09}
Tang, Y.-W., Ho, P.~T.~P., Koch, P.~M., Girart, J.~M., Lai, S.-P. \& Rao, R.
2009, \apj, 700, 251

\bibitem[2002]{t02}
Teyssier, R., 
2002, \aap, 385, 337

\bibitem[2011]{vh11}
Vlemmings, W.~H.~T., Humphreys, E.~M.~L. \& Franco-Hern\'andez, R. 
2011, \apj, 728, 149


\bibitem[1990]{wk90}
Wardle, M. \& K\"onigl, A. 
1990, \apj, 362, 120

%





\end{thebibliography}
\end{document}